\documentclass{aastex63}

\received{}
\revised{\today}
\accepted{}

\submitjournal{AASJ}

\shorttitle{Inner Moons of Uranus}
\shortauthors{\'Cuk et al.}

\begin{document}

\title{Cupid Is Not Doomed Yet: On the Stability of the Inner Moons of Uranus}
%\textcolor{red}{I would propose a title such as: Stability State in the Inner Moons of Uranus OR Stability investigation (or study) of the Inner Moons of Uranus}}
% I would go for "Cupid May Not be Doomed After All - RF

\correspondingauthor{Matija \'Cuk}
\email{mcuk@seti.org}

\author[0000-0003-1226-7960]{Matija \'Cuk}
\affiliation{SETI Institute \\
339 N Bernardo Ave\\
Mountain View, CA 94043, USA}

\author[0000-0002-9527-7920]{Robert S. French}
\affiliation{SETI Institute \\
339 N Bernardo Ave\\
Mountain View, CA 94043, USA}

\author[0000-0002-8580-4053]{Mark R. Showalter}
\affiliation{SETI Institute \\
339 N Bernardo Ave\\
Mountain View, CA 94043, USA}

\author[0000-0002-2736-3667]{Matthew S. Tiscareno}
\affiliation{SETI Institute \\
339 N Bernardo Ave\\
Mountain View, CA 94043, USA}

\author[0000-0002-4416-8011]{Maryame El Moutamid}
\affiliation{Cornell Center of Astrophysics and Planetary Sciences\\ Department of Astronomy and Carl Sagan Institute\\
Cornell University\\
326 Space Science Building\\
Ithaca, NY 14853, USA}

\begin{abstract}

Some of the small inner moons of Uranus have very closely-spaced orbits. Multiple numerical studies have found that the moons Cressida and Desdemona, within the Portia sub-group, are likely to collide in less than 100 Myr. The subsequent discovery of three new moons (Cupid, Perdita, and Mab) made the system even more crowded. In particular, it has been suggested that the Belinda group (Cupid, Belinda, and Perdita) will become unstable in as little as 10$^5$ years. Here we revisit the issue of the stability of the inner moons of Uranus using updated orbital elements and considering tidal dissipation. We find that the Belinda group can be stable on $10^8$-year timescales due to an orbital resonance between Belinda and Perdita. We find that tidal evolution cannot form the Belinda-Perdita resonance, but convergent migration could contribute to the long-term instability of the Portia group. We propose that Belinda captured Perdita into the resonance during the last episode of disruption and re-accretion among the inner moons, possibly hundreds of Myr ago.   

\end{abstract}


\section{Introduction} \label{sec:intro}

Uranus has, as of early 2022, 27 known moons. Apart from the five large moons (Miranda, Ariel, Umbriel, Titania, Oberon) known before the space age, the remaining satellites are divided into nine outer (irregular) moons that were likely captured by the planet \citep{gla98, gla00, kav04, she05}, and thirteen inner moons orbiting closer to Uranus than Miranda. Ten of the inner moons were discovered in 1986 during the Voyager 2 fly-by \citep{smi86}, Perdita was discovered in Voyager data years later \citep{kar01}, while Mab and Cupid were found using the Hubble Space Telescope \citep{sho06}. This paper will concentrate on the dynamics of the inner moons, which all have closely-packed low-eccentricity orbits close to the equatorial plane of the planet (Tab. \ref{table1}). 

% I find it really confusing that "unlabelled radii as well as elements with an asterisk" are from XXX. If you're going to mark a source, make it consistent everywhere. FIXED

\begin{deluxetable*}{lcrrrrrr}
\tablenum{1}
\tablecaption{Orbital elements and mean radii for the inner moons of Uranus. Unlabelled orbital elements are from \citet{fre17} (the epoch is 2008-09-01 00:00:00 UTC). Radii and elements with an asterisk (*) are from \citet{sho06}. Radii with a dagger ($\dagger$) are from \citet{kar01}. Double daggers ($\ddagger$) label the mean elements obtained from the Jet Propulsion Laboratory's Solar System Dynamics website on February 5, 2021.  \label{table1}}
\tablewidth{0pt}
\tablehead{
\colhead{Moon} & \colhead{Radius} & \colhead{Semimajor} & \colhead{Eccentricity} & \colhead{Inclination} & \colhead{Mean Longitude} & \colhead{Longitude of} & \colhead{Longitude of}\\
\colhead{Name} & \colhead{[km]} & \colhead{Axis $a$ [km]} &  \colhead{$e$} &
\colhead{ $i$ [$^{\circ}$]} & \colhead{ $\lambda$ [$^{\circ}$]} & \colhead{Pericenter $\varpi$ [$^{\circ}$]} & \colhead{Asc. Node $\Omega$ [$^{\circ}$]}}
%\decimalcolnumbers
\startdata 
Cordelia & $^{\dagger}$21 $\pm$ 3 & $^{\ddagger}49,800\phantom{.000}$ & $^{\ddagger}0.0003\phantom{0}$  & $^{\ddagger}0.085\phantom{00}$ & & & \\
Ophelia & $^{\dagger}$23 $\pm$ 4 & 53,763.797 & 0.00948 & 0.09080 & 231.99175 & 270.59183 & 87.00642 \\
Bianca & $^*$27 $\pm$ 2 & $^*$59,165.562 & $^*$0.00027 & $^*0.1811\phantom{0}$ & & & \\
Cressida & $^*$41 $\pm$ 2& 61,766.760 & 0.00056 & 0.05023 & 137.70117 & 274.65569 & 207.44872 \\
Desdemona & $^*$35 $\pm$ 4 &62,658.422 & 0.00073 & 0.04673 & 352.89892 & 157.93454 & 122.42866 \\
Juliet & $^*$53 $\pm$ 4 & 64,358.268 & 0.00122 & 0.03518 & 76.17466 & 89.97098 & 144.28899 \\
Portia & $^*$70 $\pm$ 4& 66,097.314 & 0.00051 &  0.01643 & 50.43991 & 6.91932 & 31.75335 \\
Rosalind & $^*$36 $\pm$ 6& 69,926.849 & 0.00090 &  0.04922 & 139.85182 & 239.72554 & 249.40149 \\
Cupid &  $^*$8.9 $\pm$ 0.7 & 74,392.338 & 0.00047 &  0.07028 & 109.06840 & 320.20439 & 59.71798 \\
Belinda & $^*$45 $\pm$ 8 & 75,255.674 & 0.00079 &  0.00172 & 324.66202 & 38.78082 & 232.28659 \\
Perdita & $^*$13.3 $\pm$ 0.7 & 76,416.764 & 0.00345 &  0.05889 & 67.48953 & 338.81201 & 248.66870 \\
Puck & $^*$81 $\pm$ 2 & 86,004.545 & 0.00010 &  0.33114 & 306.44436 & 136.73741 & 235.33913 \\
Mab & $^*$12.4 $\pm$ 0.5 & 97,735.966 & 0.00347 & 0.12217 & 153.06479 & 149.92208 & 84.63121 \\
\enddata
%\tablecomments{}
\end{deluxetable*}

The inner moons of Uranus are known to be dynamically unstable. \citet{dun97} found, using numerical simulations, that the then-known inner moons of Uranus would suffer mutual collisions on timescales between 4 and 100~Myr, much shorter than the 4.5 Gyr age of the Solar System. While subject to uncertainties of the albedos and densities (and therefore masses) of the inner moons, these simulations robustly demonstrated that this packed system is inherently unstable. Subsequent work by \citet{mey05} supported the findings of \citet{dun97}. 

As more inner moons of Uranus were discovered \citep{kar01, sho06}, the stability of the inner system became even more unlikely. \citet{fre12} revisited the long-term stability of the Uranian inner moons, including the new discoveries. They found that there are two distinct sources of instability in the system. The Belinda group, which includes recently found Cupid and Perdita, is unstable on timescales as short as $10^5$ years. On the other hand, members of the Portia group tend to be unstable on multi-Myr timescales \citep[consistent with the results of ][]{dun97}, with the satellites Cressida and Desdemona being usually the first to collide. In general, instability in the system is driven by overlap of first order resonances, often between trios of moons \citep{daw10, qui14, fre15}.

The relatively short-term instability of the Uranian inner moons implies that they have experienced repeated cycles of destruction and re-accretion since the formation of the Solar System. The presence of rings in-between moons \citep{smi86, sho06} reinforces the idea that collisions may happen periodically, generating the ring material. As successive generations of the satellite system would be different in their distribution of moons, one would expect that the system would evolve into a more stable configuration after many iterations, so having lifetimes as short as $10^5$ yr after 4.5 Gyr of evolution is somewhat counter-intuitive. Could some dynamical effects that the prior studies have overlooked make the system longer-lived?

A stable system may at first appear unstable if the phase-space of initial conditions contains very small islands of stability that are only a subset of (or overlap only a little with) the parameter range allowed by the observations. This is a common situation in studies of exoplanetary systems due to high uncertainties in the masses and eccentricities  of many known exoplanets. \citet{fab10} and \citet{dec12} both analyzed systems that were typically unstable on timescales short compared to stellar ages. When dynamical analyses found small regions of parameter space that lead to longer-term stability, these were often associated with stable mean-motion resonances. In the context of Uranian satellites, this situation is more likely to apply to the more recently discovered moons such as Cupid or Perdita, and we would then expect that the orbital solutions will become more stable as more data is available. 

%Another potential factor that may help stabilize the inner Uranian system is tidal dissipation \citep[cf. ][]{mey05}. Tides within the moons should damp their eccentricities over time, potentially increasing the system's stability. Timescales for tidal dissipation is these small moons are long, although if we assume they are "rubble piles" \citep{gol09}, dissipation timescales for some of them can be as short as $10^8$~yr, approaching the upper end of the estimates of their dynamical lifetimes. Furthermore, large librations due to elongated shapes of the satellites could greatly enhance tidal dissipation \citep{yod82, qui20}. The scaling of tidal processses makes them more likely to be relevant for the larger inner Uranians, rather than smaller moons like Cupid. 

In the following sections we will first revisit the dynamics of the Belinda group which includes the shortest-lived moons \citep{fre12}, and then separately address the long-term evolution of the system in light of the instability of the Portia group that has been confirmed by multiple researchers \citep{dun97, fre12}.

\section{Dynamics of the Belinda Group}\label{sec:belinda}

The first step in evaluating the lifetime of the inner moons of Uranus is to consider the dynamics of the Belinda group, consisting of the eponymous larger moon flanked by smaller Cupid and Perdita. \citet{fre12} found that this sub-group is unstable on $10^5$~yr timescales, with Cupid usually being the first moon to suffer a collision. However, since Cupid had a much shorter observed arc at the time, its orbit and the resulting dynamics were still uncertain.

Here we use orbital solutions previously presented by \citet{fre17}. The orbital solutions explicitly include orbital precession and we treat them as geometric orbital elements, using the approach of \citet{ren06} to convert them into position and velocity vectors. We use the symplectic integrator SWIFT-rmvs4 from the SWIFT family \citep{lev94} with Uranus as the central body. In this section we will exclude all bodies other than the three members of the Belinda group. This is done to enable $>10$~Myr simulations within a reasonable amount of real time. As the interaction between these three moons is the strongest driver of chaos, we believe that our three-moon simulations still capture the most important dynamics of the group. 

%\textcolor{red}{Is this assumption justified? If so, may be add a sentence or two explaining that excluding all (but these 3 moons) would not affect the dynamics of the Belinda group, at least in this stage of the study  RF: Isn't that what the final sentence of this paragraph says?}.

The moons' densities and therefore their masses are unknown. For our ``nominal'' case, we decided to use 5:1 and 3:1 radius ratios between Belinda and Cupid, and Belinda and Perdita, respectively, to set mass ratios of 125:1 (Belinda-Cupid) and 27:1 (Belinda:Perdita). The latter ratio is based on the observed libration of Perdita \citep{fre17}. The semi-axes of irregularly-shaped Belinda are given by \citet{kar01} as $64 \times 32$~km. If we assume that Belinda is a prolate ellipsoid with the third radius also being $32$~km, the volumetric radius will be $(a b c)^{1/3} \approx 40$~km. Density is not constrained by the observations, with the lower limit set by the Roche density of $390$~kg~m$^{-3}$ \citep[][using $\gamma=1.6$]{tis13}. The only inner moon of Uranus with a measured density is Cressida, with $\rho=860 \pm 170$~kg~m$^{-3}$ \citep{cha17}, but Cressida is also significantly closer to Uranus, experiences stronger tidal forces, and may therefore have higher density than  Belinda \citep{tis13}. Within these limits we choose 600~kg~m$^{-3}$ as the ``nominal'' assumed density of Belinda group moons. This density would imply significant porosity, but is generally in line with the densities of Saturnian ring-moons \citep{tis13}. This is gives us ``nominal'' mass of Belinda $m_B=1.6 \times 10^{17}$~kg, with the nominal masses of Cupid and Perdita being 125 and 27 times smaller, respectively. 

\begin{figure*}
\epsscale{1.2}
\plotone{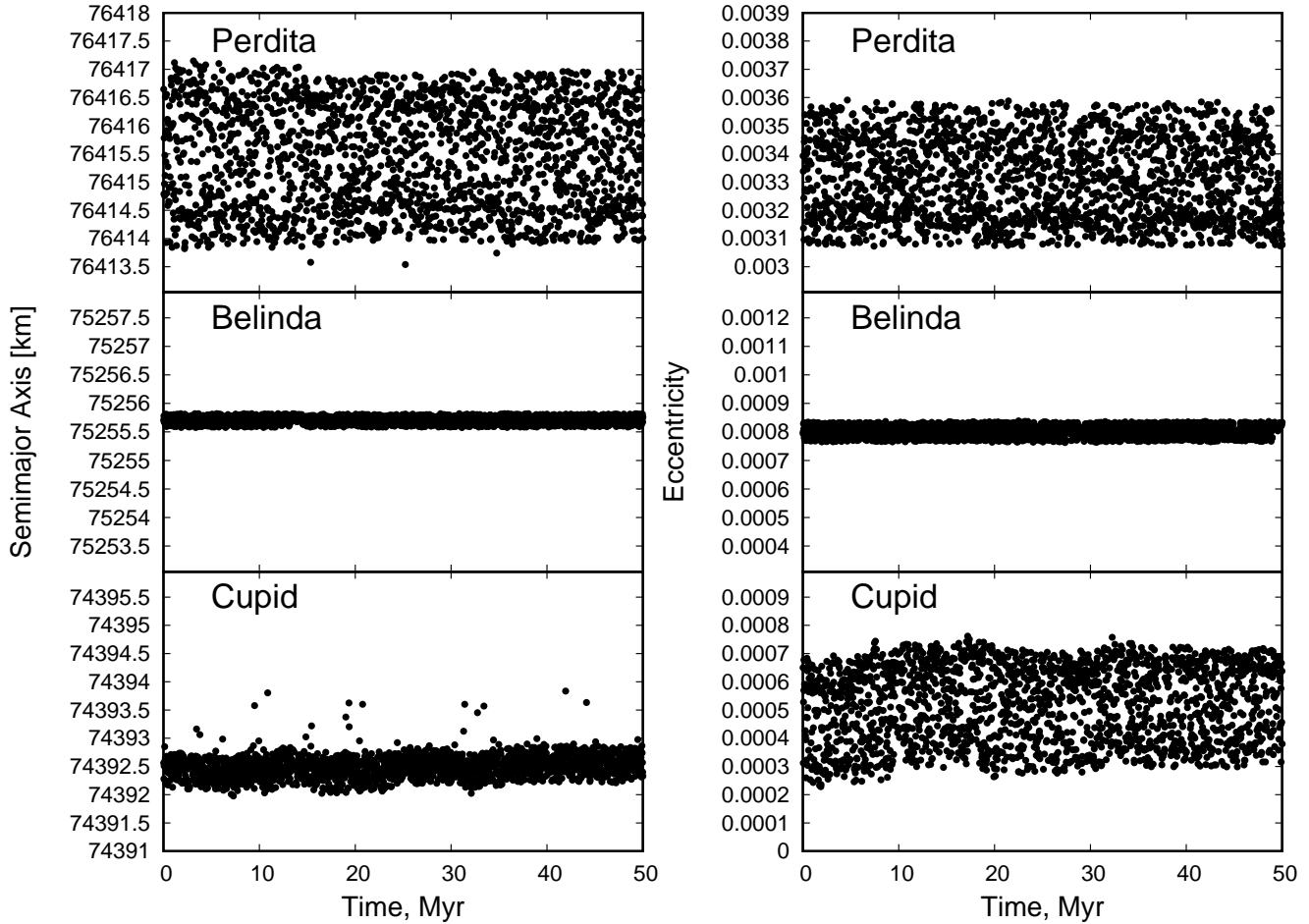}
\caption{A 50-Myr ``nominal'' simulation of the Belinda group moons using SWIFT-rmvs4 integrator and orbital elements found by \citet{fre17}. Belinda was assumed to have a mass of $1.6 \times 10^{17}$~kg, and Belinda/Cupid and Belinda/Perdita mass ratios were set to 125 and 27, respectively. The moons are stable over the whole course of the simulation.\label{nom}}
\end{figure*}

\begin{figure*}
\epsscale{.6}
\plotone{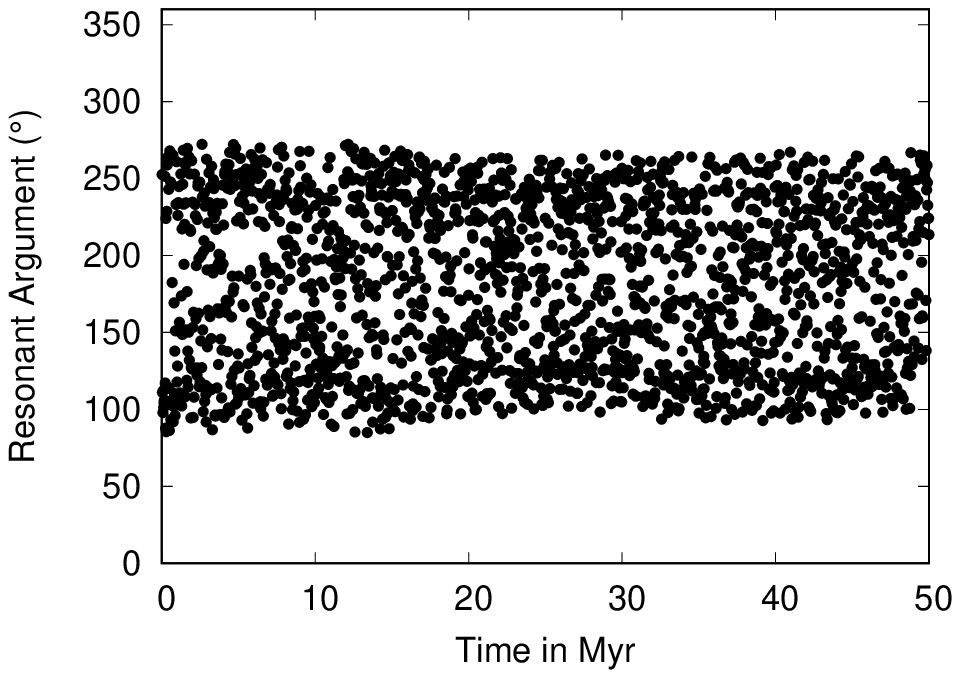}
\caption{The resonant argument $44 \lambda_P - 43 \lambda_B - \varpi_P$ of the Belinda-Perdita 44:43 MMR in the same simulation shown in Fig. \ref{nom} (variables have the same meaning as in Table \ref{table1}). The librating argument establishes that the two moons are in the resonance for the whole duration of the simulation. \label{nom_arg}}
\end{figure*}

Fig. \ref{nom} plots the outcome of the 50-Myr simulation of Cupid, Belinda, and Perdita with the ``nominal'' masses described above and orbits reported by \citet{fre17}, using SWIFT-rmvs4 with a timestep of $5 \times 10^{-5}$~yr (0.0183 days). The trio of moons is stable over 50 Myr, and while the orbits of Belinda and Perdita appear rather regular, the orbit of Cupid shows some signs of chaos. The orbits of Belinda and Perdita are stabilized by their mutual 44:43 mean-motion resonance (MMR), the resonant argument of which is plotted in Fig. \ref{nom_arg}. This long-term stable resonance between Belinda and Perdita is a direct result of using the updated orbits fits published by \citet{fre17}. \citet{sho06} and \citet{fre17} already identified the Belinda-Perdita 44:43 MMR as necessary to fit the observations, but did not explore the resonance's implications for the system's long-term stability. All previous orbital solutions place these two moons outside of the long-term stable region of the resonance, resulting in collisions (usually between Belinda and Cupid) on $10^5$ year timescales \citep{fre12}. 

\begin{figure*}
\epsscale{1.2}
\plotone{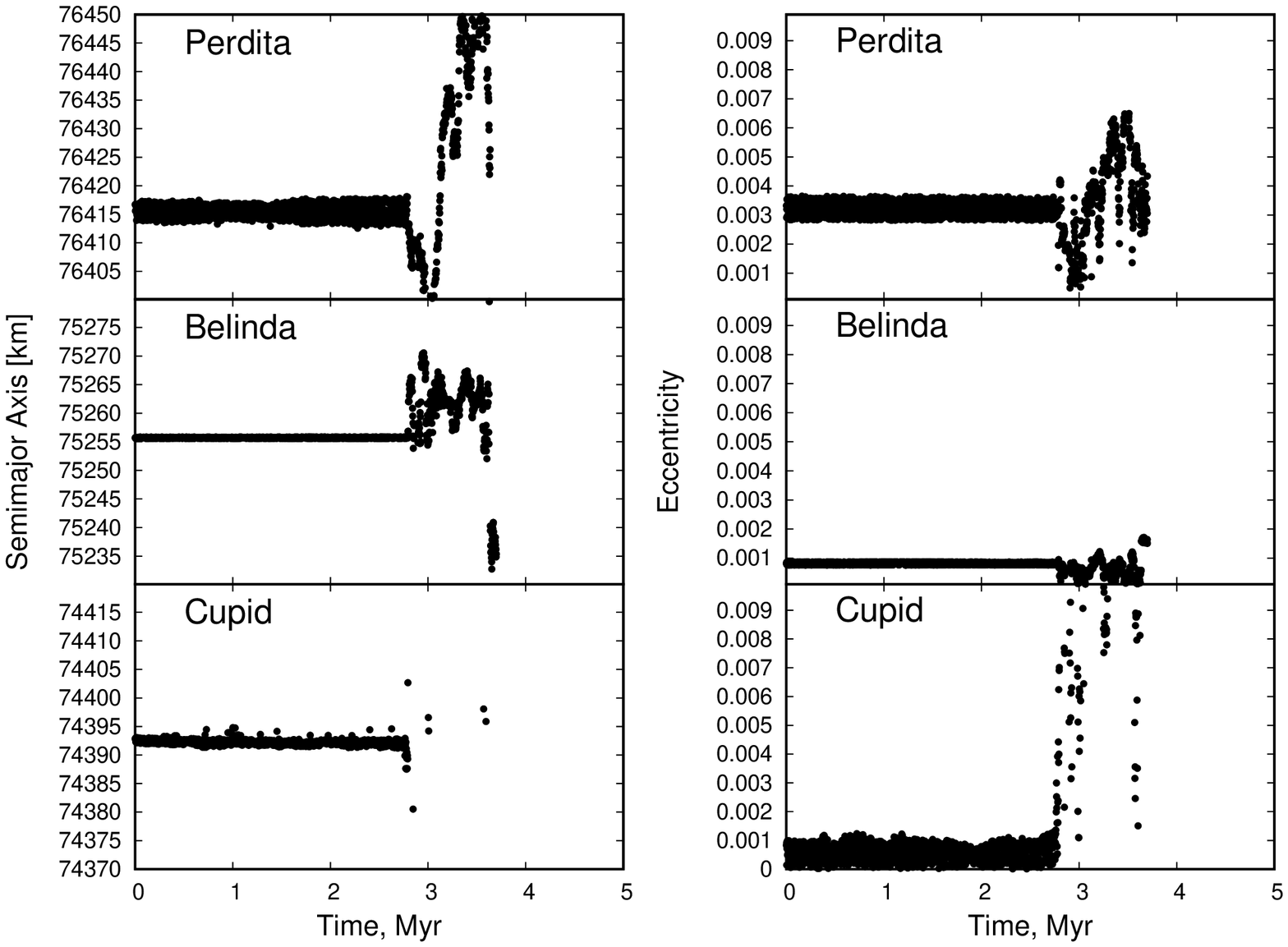}
\caption{Simulation identical to the one shown in Fig. \ref{nom}, except that the masses of all three moons have been increased by a factor of 2. The simulation results in an instability in less than 3~Myr.\label{inst}}
\end{figure*}

To show that the stability of the Belinda group is dependent on the masses of the satellites, in Fig. \ref{inst} we plot a simulation with the same initial conditions as in Fig. \ref{nom}, but with twice larger masses for all three Belinda group satellites. The moons become unstable in less than 3 Myr, indicating that the Belinda group members must have a density lower than 1200 kg~m$^{-3}$, the most likely density of Miranda.

In order to further explore the dependence of the stability on satellite masses, we ran an additional eight 50 Myr integrations in which we varied the masses of Belinda and Perdita (while the mass of Cupid was kept the same as in Fig. \ref{nom}). In these simulations the mass of Belinda and Perdita was set at either 2/3, 1, or 3/2 of the value used in the ``nominal'' simulation (Fig. \ref{nom}). Together with the ``nominal'' simulation, these eight runs make a complete 3x3 grid in the assumed masses of Belinda and Perdita. Seven of the simulations stayed stable until 50 Myr, indicating that a wide range of densities $\leq 900$~kg~m$^{-3}$ are compatible with the stable system. The only unstable simulation in this set was one with the nominal mass of Belinda and $50\%$ larger mass of Perdita, which went unstable after about 40~Myr. This suggests that the Belinda/Perdita mass ratio is also relevant to the stability of the system. While the satellite masses are largely unconstrained, \citet{fre17} found from observations that the Belinda/Perdita mass ratio is about 27, so we have reason to think that our ``nominal" case (as well as the two other 50-Myr stable simulations that have the same mass ratio) should be more representative of the real system, strengthening the case for stability. 

%\textcolor{red}{What does "relative stability" mean? Relative to what?} MC: relative to previous results (10^5 year), although we don't know about 5 Gyr

We conclude that the Belinda group is likely to be relatively stable, at least on 50 Myr timescales. There are several caveats to this conclusion. One is that we have not included satellites other than Cupid, Belinda, and Perdita in our simulations, as adding in more moons would make the time of the simulation using our present software and hardware impractical.  As the inner moons outside Belinda group are well separated from the group and the dynamics is dominated by Uranus's oblateness, we think that the stability is likely to hold with the other moons' perturbations included. 

%However, Belinda group may be affected by an instability that started elsewhere among the inner moons. Once we exclude the Belinda group, the next least stable moon pair are Cressida and Desdemona, within the Portia group \citep{dun97}, which we address in the later Sections.

\section{Tidal Evolution of the Inner Moons}\label{sec:uranus}

The existence of a stable orbital resonance between Belinda and Perdita requires a mechanism for the formation of this resonance. In general, capture into mean-motion resonances requires convergent orbital migration of the two bodies \citep{md99}. Therefore, either Belinda migrated outward or Perdita migrated inward. Currently, the most significant orbital evolution mechanism acting on the moons is tidal dissipation within Uranus, and in this section we will examine its likely effects.

\begin{figure*}
\epsscale{1}
\plotone{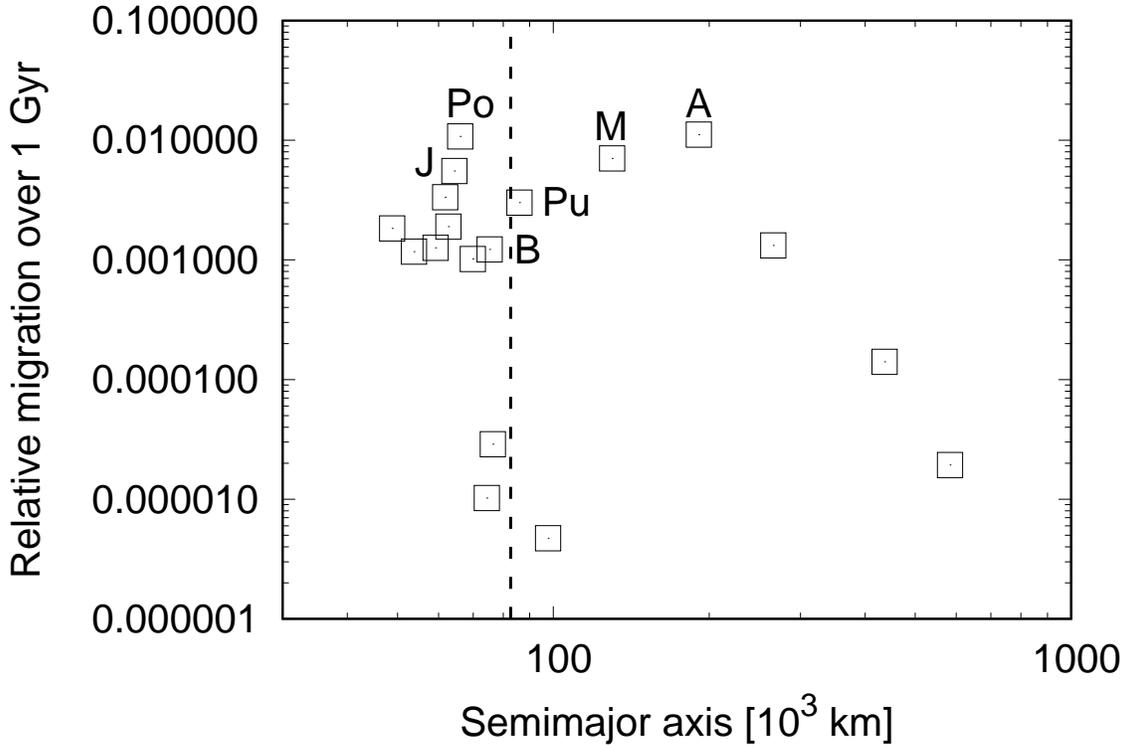}
\caption{Relative amount of orbital evolution $|\Delta{a}/a|$ over 1 Gyr for the regular moons of Uranus, assuming Love number $k_2=0.1$ and tidal quality factor $Q=15,000$ for Uranus. Known masses have been used for the five large satellites \citep{cuk20}, while we used radii from Table \ref{table1} and density of $850$~kg~m$^{-3}$ for the inner moons. Labelled moons are Juliet (J), Portia (Po), Belinda (B), Puck (Pu), Miranda (M), and Ariel (A). The dashed line plots the distance to the synchronous orbit from the planet's center; the moons interior to it migrate inward, while the moons exterior to this line migrate outward.\label{tidal}}
\end{figure*}

In the classical theory of tidal evolution, a satellite raising tides on the planet will migrate at the rate \citep{md99}:
\begin{equation}
\dot{a}= {\rm sign}(\Omega_p - n) {3 k_2 \over Q}{m_s \over m_p} \left({R \over a}\right)^5 n a
    \label{adot}
\end{equation}
where $a$, $n$, and $m_s$ are respectively the satellite's semimajor axis, mean motion, and mass, while $\Omega_p$, $k_2$, $Q$, $m_p$, and $R$ are respectively the planet's rotation rate, tidal Love number, tidal quality factor, mass, and radius. Using Eq. \ref{adot} we plot in Fig. \ref{tidal} $|\Delta a/a|$ over 1 Gyr for all regular moons of Uranus, assuming a frequency-independent tidal quality factor $Q=15,000$ for Uranus \citep{cuk20}. 

An important feature of Eq. \ref{adot} is that moons interior to the synchronous orbit (which is at about $a=82,700$~km) migrate inward, toward Uranus. This applies to all the inner moons except for Puck and Mab (with the latter's tidal migration being negligible due to low mass). Since Belinda is interior to the synchronous orbit, it should migrate inward much faster than the less-massive Perdita, making their orbits diverge. Divergent migration pushes two moons out of, rather than into, their mean-motion resonance. Therefore the origin of the observed Belinda-Perdita 44:33 resonance requires an explanation different from tidal evolution.

While the tidal evolution of Belinda cannot explain its resonance with Perdita, it can be used to constrain the dynamical age of Cupid. Cupid is only about 4~km exterior to the 58:57 resonance with Belinda, and first-order resonances with Belinda were shown by \citet{fre12} to be a major source of instability for Cupid. Given that Belinda is estimated to migrate inward by 0.1\% of its semimajor axis (or about 75 km) over 1~Gyr, its interior 58:57 resonance should have encountered Cupid about 50~Myr ago. This estimate assumes a frequency independent $Q$. A ``constant-lag'' approach \citep[cf.][]{can99} would estimate an effective tidal $Q$ of Uranus at Belinda's orbital frequency $Q_B= Q_A (n_B-\Omega_p)/(\Omega_p-n_A)$, where subscripts A and B refer to Ariel and Belinda \citep[as the estimate of tidal $Q$ by ][ is derived from the dynamics of Ariel]{cuk20}. This gives us $Q_B \simeq 4.7 Q_A$, meaning that the lifetime of Cupid could be as long as $200-250$~Myr depending on the nature of tidal dissipation within Uranus. Coupled with the wide range of Belinda's mass that would result in a stable resonance with Perdita (as shown in the previous Section), the maximum allowed dynamical age of Cupid is about 0.5~Gyr. Despite large uncertainties, we can say that Cupid is almost certainly not primordial, with potential implications for the age of the Belinda-Perdita resonance. 

Figure \ref{tidal} also shows that Portia, the largest moon of the unstable Portia group \citep{dun97, fre12} should migrate by about 1\% of its orbital distance over 1~Gyr, which is comparable with the fastest-evolving large moon, Ariel. Given that the rate of orbital evolution is proportional to mass, Portia should migrate inward faster than Juliet, which should in turn migrate faster than Desdemona. The end result is that the Portia group becomes more compact over time, likely to the detriment of its dynamical stability. Apart from the direct effect of the smaller separation between orbits, this relative migration is likely to result in resonant encounters like the ones discussed above for the Belinda group, adding to dynamical excitation. The separations and masses within the Portia group generally result in chaotic mean-motion resonances. When mean-motion resonances are encountered, the likely outcome is dynamical excitation rather than stable capture.

Constant tidal evolution resulting in the tightening of orbital separations may be the reason why the Portia group appears to be unstable on $<100$~Myr timescales \citep{dun97}, even though the planet is billions of years old. Moons are thought to form from disks or rings, and typically the number and separation of objects that are observed in mature systems is controlled by their stability. However, even a modest amount of long-term tidal evolution may be sufficient to push a previously stable system ``over the edge'' into an unstable state. 

%\textcolor{red}{OK so you're claiming that we are looking at the Uranian system at a special time, where we just happen to be near the end of life of the Portia group?} MC: That was the idea, now I think maybe we are seeing the time before Cressida-Desdemona merger?

Apart from tides raised by the moons on Uranus, the planet will also raise tides on the satellites. As we expect the inner moons to be in synchronous rotation, the main effect of tidal dissipation within them is damping of their eccentricities. In case of very elongated satellites rotational librations may be greatly enhanced over the direct geometric effect of the orbital eccentricity \citep{tis09}. These rotational librations also contribute to eccentricity damping, and may be the dominant contribution \citep{yod82, efr18, qui20}. 

We performed some simulations exploring the effect of satellite tides on the stability of the Portia group and found that even greatly enhanced dissipation could not stabilize the Portia group in a steady-state comparable to the present configuration. The orbits of the Portia group members are currently excited, with much of the angular momentum deficit \citep{las97} contained in the relatively large eccentricity of Juliet ($e=0.0012$)\footnote{Much smaller eccentricity of Juliet reported by \citet{sho06} has since been found by R.S.~French to be an artifact of orbit fitting; re-analysis of the pre-2006 data finds they are best fitted with $e \approx 0.001$ for Juliet}. Juliet's eccentricity can be communicated to Desdemona and Cressida through angular momentum exchange, eventually leading to orbit crossing. Therefore, we can only confirm previous findings by \citet{dun97} and \citet{fre12} on the instability of the Portia group. The fact that the system could be more long-lived if Juliet had a more circular orbit suggests that the actual age of the Portia group could be significantly longer than its remaining lifetime ($<100$~Myr). 

\section{Other Mechanisms of Orbital Evolution}

In order to evaluate different migration mechanisms that could have lead to the formation and evolution of the resonance, it is helpful to calculate the amount of migration necessary to form the Belinda-Perdita resonance. Assuming that migration is restricted to Belinda and that Perdita's pre-resonance eccentricity was zero, we get \citep[using the approach from Section 8.13 in][]{md99}:
\begin{equation}
\Delta a_B \approx 44 e^2_P a_B  
\end{equation}
where the subscripts B and P refer to Belinda and Perdita, and the factor of 44 is due to this being a 44:43 MMR. Given that $e_P=0.0035$, the amount of migration by Belinda necessary to form the resonance is about 40~km. This is the minimum amount of relative migration as Belinda may have had moved inward due to tides since the formation of the resonance, in which case Perdita was initially deeper in the resonance. Even if Cupid formed at approximately the same epoch as the Belinda-Perdita resonance was established, it must have assumed its current orbit only once the evolution of the Belinda-Perdita resonance was over, or otherwise Cupid would have encountered multiple resonances with Belinda.

Tidal evolution is not the only orbital migration mechanism available to planetary moons. If the planet has rings interior to the moons, ring-moon interactions will generally lead to an outward migration of the moon \citep{gol82, hed18}. The rings of Uranus are currently relatively thin and narrow, but they could not have lasted in the present configuration for more than about $6 \times 10^8$~yr \citep{esp89}. In particular, if the Portia group suffers periodic instabilities, during which the moons collide forming a ring which then re-accretes into satellites, there may have at times existed a massive ring interior to Belinda's orbit. This idea of a periodic ring-moon cycle is reminiscent of the model of a past Martian ring-moon system proposed by \citet{hes17}, except that, at Uranus, the system of multiple moons would be destroyed by mutual collisions rather than tidal forces. 

Alternatively, the Portia group may follow the \citet{hes17} hypothesis even more closely, and represent the latest iteration of the ring-moon cycle in which the moons are destroyed by tidally evolving inward, beyond the Roche limit. In this scenario, a previous generation of inner moons in the location of the Portia group were more massive, experienced faster inward tidal evolution, and were tidally disrupted once they reached their Roche limit. The resulting ring would have been Roche-interior and therefore long-lived, and over time it produced Portia and its fellow group members from its outer edge, as has been proposed for the inner moons of Saturn \citep{cha10}. \citet{hes17} find that each cycle of disruption and re-accretion at Mars produces a moon about five times less massive than the previous one; we would expect a similar ratio in mass between the proto-Portia and the current Portia group.

The mass of Portia is not known, but if we assume a mean radius of 70~km and a density of 860~kg~m$^{-3}$ \citep[like Cressida;][]{cha17}, we can estimate it to be $1.2 \times 10^{18}$~kg. This is about an order of magnitude less massive than the rings of Saturn \citep{ies19}. However, given the smaller scale of the Uranian system, if all four major members of the Portia group were to be thoroughly disrupted and spread into a ring, the surface density of this ring would be comparable to the main rings of Saturn. This ring would not be as long-lived as those of Saturn, as the Portia group is demonstratively beyond the relevant Roche limit, and is likely to re-accrete into moons as soon as the collisions damp the dynamical excitation of debris. On the other hand, if the more massive precursor of the Portia group were to evolve within the Roche limit, the resulting ring would be long-lived, and would likely spread over a large range of radial distances.

%\textcolor{red}{I don't see why we're talking about the mass and surface density of Saturn's rings here. Is there something special about those rings that makes them relevant?} MC: To give people idea about the density and appearance

If there was an epoch when the precursors to the Portia group were reduced to a ring (either through collisions or tidal disruption), other moons would have reacted gravitationally with that ring. Interaction is strongest for the moons with low-order resonances in the ring \citep{gol79, taj17}. Rosalind, which is about 4,000~km outside the orbit of Portia, may or may not have formed together with the Portia group. If it did not, and was there to interact with the ring, Rosalind would have experienced outward evolution due to interactions with the ring though numerous resonances. Much larger Puck has a 3:2 inner MMR between the current orbits of Portia and Juliet and a 2:1 MMR close to Ophelia, so it would interact with a temporary ring at the present location of the Portia group, but would not have any 1st-order resonances in a longer-lived Roche-interior ring.  As Puck is about as massive as the Portia group, the amount of orbital evolution Puck can experience through interactions with the ring would be limited. Cupid, Perdita, and Mab, if they existed at the time, would have much weaker interactions with the ring due to their small masses.

This leaves us Belinda, which is a sizeable moon that would have first order resonances in the proto-Portia-group ring. The 5:4 and 6:5 inner MMRs with Belinda are just outside the orbits of Juliet and Portia, respectively, so Belinda would likely interact with a ring at the current location of the Portia group. Also, the 2:1 inner MMR with Belinda is interior to Cordelia, within the present ring system, making interactions with Roche-interior rings possible. If Belinda migrated outward due to interactions with the past ring that gave rise to the Portia group, this migration could potentially explain the Belinda-Perdita MMR.

We can estimate the torque due to a first-order $(p+1):p$ MMR using the following expression \citep{taj17}:
\begin{equation}
T_R \approx 8.6 p^2 \left({m_s \over m_p} \right)^2 \Sigma a^4 n^2
\end{equation}
Where $p$ is an integer and $\Sigma$ is the ring surface density at the location of the resonance. We can use this expression to find the rate of migration:
\begin{equation}
{\dot{a} \over a} = 2 {\dot{L} \over L} \approx 17.2 p^2 {\Sigma a^2 \over m_p} {m_s \over m_p} n    
\end{equation}
where $L=m_s a^2 n$ is the satellite's angular momentum. Assuming $\Sigma = 1000$~kg~m$^{-2}$, we get $a/\dot{a} \approx$ 4~Gyr for the 6:5 MMR between the ring (at the orbit of Portia) and Belinda. In order for Belinda to move the 40~km required to evolve the resonance with Perdita, it would need to interact with the Roche-exterior ring for about 2~Myr. This is not plausible, as we would expect the Perdita group to re-accrete on much shorter timescales. Evolution of the Belinda-Perdita resonance through the 2:1 MMR between a Roche-interior ring and Belinda would take tens of Myr, but this process is more plausible as the Roche-interior ring would be more long-lived. Therefore, if the Belinda-Perdita resonance was assembled by ring-torques on Belinda, a long-lived, Roche-interior ring is needed.

We tentatively conclude that a ring-moon cycle similar to the one \citet{hes17} have proposed for Mars may have been operating at Uranus. \citet{hes19} already did propose that ring-moon cycles operate at Uranus, but they concentrated on the history of Miranda, which is arguably too far from the planet to interact with the ring. While the exact position of the ring's outer edge depends on cohesive forces between ring particles, Belinda's inner 2:1 MMR is expected to be within any past massive rings. The torques arising from Belinda-ring interaction would drive outward migration of Belinda, explaining its current 44:43 MMR with Perdita. In this model, the Portia group and its progenitors would be in the ``boomerang'' regime of ring-moon evolution, as defined by \citet{hes19}, in which the moons first evolve outward by ring torques and then migrate inward across the Roche limit due to tides.

A possible alternative to the migration of Belinda would be orbital evolution of Perdita due to a drag force. A force proportional to the surface area would affect Perdita more than Belinda, possibly explaining their resonance. Gas drag was likely a factor only during the initial formation of the system, 4.5~Gyr ago. If the resonance was assembled by gas drag the MMR would also need to be primordial, which we cannot strictly exclude but appears unlikely in the light of limited stability of the inner moons (and this scenario would require Cupid to be a much later addition). 

In the present system the most important drag force on Perdita is the radiative Binary Yarkovsky-Radzievskii-O'Keefe-Paddack (BYORP) effect \citep{cuk05}. The BYORP effect makes irregularly-shaped synchronously-rotating moons drift in semimajor axis due to asymmetry in thermal radiation from their leading and trailing hemispheres. The direction of the drift is random and depends only on the shape of the satellite. \citet{cuk05} found that the orbit of a typical Near-Earth Asteroid (NEA) binary at 1~AU with a $R=150$~m secondary evolves with a timescale of about $3 \times 10^4$~yr. We can estimate the timescale for Perdita's orbital evolution as:  
\begin{equation}
t_P = {v \over \dot{v}} \approx 3 \times 10^{4} {\rm yr} \left({v \over 0.1~{\rm m/s}}\right) \left({R \over 150~{\rm m}}\right) \left({a_U \over 1~{\rm AU}}\right)^2  
\end{equation}
as the BYORP acceleration is proportional to the incident solar flux and the moon's surface area, and inversely proportional to the satellite's mass. The flux at Uranus is $\approx 400$ times weaker than at Earth, and Perdita orbits about $10^5$ times faster and is 100 times larger than our typical NEA secondary, so we estimate the timescale for its BYORP evolution to be $\approx 10^{14}$~yr. Over the age of the solar system Perdita is likely to migrate by at most $\approx 4$~km (in an unknown direction), making Perdita's BYORP-driven migration an order of magnitude less important than our (lowest) estimate for the tidal evolution of Belinda. Therefore we can exclude BYORP as a mechanism for the formation of the Belinda-Perdita resonance.     

\section{Satellite Lifetime Against Impacts}\label{sec:impact}

Small satellites close to giant planets experience enhanced bombardment by heliocentric impactors such as comets and Centaurs, both due to gravitational focusing by the planet and high relative velocities near the planet \citep{zah03}. Even some of the larger moons may have suffered disruption during the high-bombardment periods early in the Solar System's history \citep{nim12, mov15, won19}. Even after the end of the heavy bombardment, it is likely that collisional disruption of small ring-moons by cometary/Centaur bombardment was an important process over the age of the solar system \citep{col92, col93, col00}.

The study of \citet{zah03} was the most thorough examination to date of cratering on outer planet satellites in the current planetary architecture. At the time of their study there were large uncertainties concerning the size distribution of heliocentric impactors in the outer Solar System. More recent results from the Pluto-Charon system established that the \citet{zah03} ``Case A'' model impactor distribution largely matches the Kuiper Belt-derived impactors \citep{sin19}, while the steeper ``Case B'' size-distribution (seen at Neptune and Saturn) is likely to be planetocentric in origin \citep{kir22}. Crater size-distribution on Uranian moons generally match those on Charon \citep{kir22}, and are therefore likely to be heliocentric. We are specifically interested in a steady-state heliocentric impactor population here (as planetocentric bombardment is likely to be highly episodic), and we will assume it corresponds to the \citet{zah03} ``Case A'' model population. 

\citet{zah03}, using their ``Case A'' model, find the lifetime against impacts of all Portia-group moons to be about 2~Gyr, much longer than their dynamical lifetime. They also estimate that the lifetime of Belinda is 2.6~Gyr; Perdita was not included in this study, while Cupid was not yet discovered at the time. We can estimate the lifetimes of the small moons by noting that their binding energy $\simeq G m^2 R^{-1}$ is proportional to $R^5$, while the energy of the impactor is proportional to $R_i^3$ (velocities can be considered the same as for impacts on all Belinda-group moons). Consequently, the radii of the smallest disrupting impactor and the target moon are related as $R_i \propto R^{5/3}$. Therefore, a moon five times smaller in radius (as Cupid is relative to Belinda) can be disrupted by a $5^{5/3}=14.6$ times smaller impactor.

The (cumulative) size-distribution of small impactors appears to be close to 
$N_i (> R_i) \propto R_i^{-1}$ \citep{sin19}. Therefore, there would be about 15 times more impactors capable of disrupting Cupid than Belinda. However, the cross-section of Cupid is about 25~times smaller (we can assume similar gravitational focusing by Uranus), making Cupid paradoxically longer-lived against impacts than Belinda! While this may appear counter-intuitive, this is a natural consequence of a shallow size-distribution of small KBO-derived impactors and agrees with the results of \citet{zah03}, who find similar lifetimes against collision for all ring-moons of Uranus (assuming the ``Case A'' model). The collisional lifetime of Perdita should be comparable to that of Cupid.

These estimated collisional lifetimes of the inner moons of Uranus suggest that disruption by cometary and Centaur impacts is likely to be much less frequent than mutual collisions due to dynamical instabilities. 

\section{Conclusions}\label{sec:cons}

Our conclusions can be summarized as follows:

1. Using orbital elements determined by \citet{fre17}, we show that the Belinda group of Uranian satellites (including Belinda, Perdita, and Cupid) can be stable on 50 Myr timescales, resolving the issue of these moons' extremely short lifetimes found by \citet{fre12}.

2. The stability of the Belinda group is maintained due to a 44:43 mean-motion resonance between Belinda and Perdita, with a resonant argument of $44 \lambda_P - 43 \lambda_B - \varpi_P$.

3. The establishment of the Belinda-Perdita resonance requires past convergent smooth orbital evolution of at least 40~km. This direction of orbital migration cannot be explained by tidal evolution.

4. We propose that the resonance was established by outward migration of Belinda due to interactions with a past ring of Uranus. This ring would be a product of past tidal disruption of the progenitor of the Portia group of Uranian satellites, as a part of a ring-moon cycle proposed by \citet{hes17, hes19}. 

5. We find that the Portia group members experience significant tidal evolution which largely acts to move the moons closer to each other, potentially explaining their continuing instability this late in the history of the Uranian system. 

Our conclusions should be useful for the future work on the Uranian system, including a possible spacecraft mission. Future spacecraft data would help support or falsify our inferences about the long-term dynamical evolution of the Uranian system.
\acknowledgments

M\'C and MEM are supported by NASA Solar System Workings Program award 80NSSC19K0544. The authors thank an anonymous reviewer who helped improve the manuscript.

\bibliography{refs}{}
\bibliographystyle{aasjournal}

\end{document}